\documentclass[11pt,a4paper]{article}
\usepackage{jheppub_kim}
\usepackage{longtable}

\usepackage{epsfig}
\usepackage{amsmath}
\usepackage{graphicx}
 \usepackage{graphics}
\usepackage[hang,nooneline,scriptsize]{subfigure}

\usepackage{array}
\begin{document}

\title{Non-perturbative  correction to  thermodynamics of  conformally dressed $ 3D$  black hole }

\author[a]{Saheb Soroushfar \footnote{Corresponding Author},}
\author[b]{Hoda Farahani,}
\author[c,b]{and Sudhaker Upadhyay \footnote{Visiting Associate, Inter-University Centre for Astronomy and Astrophysics (IUCAA) Pune-411007, Maharashtra, India}}
\affiliation[a]{Faculty of Technology and Mining, Yasouj University, Choram 75761-59836, Iran}
\affiliation[b] {School of Physics, Damghan University, Damghan, 3671641167, Iran}
\affiliation[c] {Department of Physics, K.L.S. College, Magadh University, Nawada-805110, Bihar,  India}

\emailAdd{soroush@yu.ac.ir}
\emailAdd{h.farahani@umz.ac.ir}
\emailAdd{sudhakerupadhyay@gmail.com}

\abstract{We extend the study of corrected thermodynamics for the $3D$ black holes conformally coupled to scalar field up to non-perturbative level. We calculate the exponential correction to entropy  arises due to the microstate counting  for quantum states   on the boundary. This exponential correction in entropy attributes to the other thermodynamical quantities also. We study the stability and phase transition for this system of black hole under the influence of non-perturbative correction. We also discuss the quantum work associated with exponential corrected entropy. Finally, we justify the results from the view point of thermodynamic geometry. }


\maketitle

\section{Motivation and literature survey}\label{sec1}
People became interested to know about the microscopic structure (degrees of freedom)
of black holes  since the   black hole thermodynamics came to the picture \cite{1,2,3,4}.  This is because
  a black hole exhibiting Hawking temperature of  must possess its own
microscopic structure.
 Various efforts have been made in the emphasis of microscopic aspects in the holographic setting \cite{5,6,7}.
The counting
the   natural microstates of an appropriate  quantum states  representation  of a  black hole horizon in equilibrium   reproduces  correctly the
semi-classical result of entropy in an exponential form of quantum correction \cite{AA}. This form of correction  differs from the  well-known
logarithmic corrections \cite{su1,su2,su3,su4,su5,su6,su7,su8,su9,su10,su11,su12,su13} appearing in other counting schemes. This results are supposed to be significant to the Planckian regime of area.

On the other hand, the differential geometry oriented concepts have been implemented in thermodynamics of black holes aggressively. For instance,
the Ruppeiner metric \cite{ru}  (conformally) equivalent to
Weinhold's metric  \cite{we} is implemented  to describe the concept
of thermodynamic length.  These two metrics due to their phase space and the metric structures do not remain invariant under
Legendre transformations. The thermodynamics of black holes from the
thermodynamic geometry points of view  has been discussed extensively but
the exponentially corrected thermodynamics under such scheme remains untouched.

The black holes in $3D$ \cite{bt1,bt2,bt3,bt4} is quiet important to understand the basics of classical and quantum gravity.  Also, in Ref. \cite{8}, $3D$
black holes dressed with generalized dilaton gravity is discussed.
$3D$ black holes  with nonlinear electrodynamics following the weak
energy conditions is presented in Ref. \cite{9}. The $3D$ AdS black hole conformally coupled with scalar and vector fields is analyzed in \cite{10}. A  charged
black holes is dressed (non-)minimally with the scalar hair in $3D$ \cite{11,12}. It should be mention that a $3D$ general relativity is a topological theory, that the mass of the BTZ black hole is not a real gravitational mass, since the Weyl tensor which measures the free gravitational field is zero by definition in $3D$.

Here, with the intention to study the non-perturbative quantum correction to the thermodynamics of the $3D$ black hole conformally coupled with  the
 scalar field, we first revisit the action describing the model and their field equations. The theory under study does not break the conformal invariance \cite{R1}. However, the conformal invariance may be broken depending on the given theory. For example, Maxwell electrodynamics will break the conformal invariance \cite{12}. Similar theories where a length scale breaks the conformal invariance have also been discussed by Refs. \cite{R2, R3, R4}.

From the metric function, it is easy to obtain
Hawking temperature.  The equilibrium entropy can also be calculated from the Wald's formula. For the given expression of temperature and entropy, we calculate the exponentially corrected entropy which becomes significant at the quantum scale of black holes. It should be noted that the theory at hand is scale invariant, so we cannot somehow probe different energy scales with this theory. As a result this theory can only be considered as a toy model to discuss the exponential corrections to the entropy. We further derived the exponentially corrected heat capacity and found that the heat capacity for the initial state without quantum correction is completely positive and has no phase transition. This state is also established for the  quantum correction with the negative correction parameter. However, the quantum corrected heat capacity of this black hole  with correction parameter greater than  the specific value  has a zero point and has a phase transition. The small sized black holes are unstable for positive correction parameter and   correction terms with positive correction parameter do not
change the stability of black hole. Subsequently, we calculate  Helmholtz free energy and internal energy under the influence of
non-perturbative correction. Here, we observe that the exponential
correction plays a significant role to the thermodynamics of small sized
black holes.
We also shed lights on the thermodynamic geometry of the black hole. We first  calculate  the mass of
the black hole appeared in the quantum corrected effective Ruppeiner metric.
From the graphical analysis, we find that the Ricci scalar of Ruppiner metric possesses singularities.
This justifies that the divergence of the scalar curvature of Ruppiner
formalism corresponds to the phase transition point of the black hole.

The paper is outlined as following. In paper Sec. \ref{sec2}, we discuss the
metric  of  conformally dressed black holes in $3D$. We extend the study of
thermal correction for the given black hole solution to non-perturbative level in Sec. \ref{sec3}. In Sec. \ref{sec04}, we
discuss quantum work  related to the exponential corrected entropy.  In Sec. \ref{sec4}, we justify the stability and phase transition from  the perspectives of thermodynamic geometry.
Finally, we summarize results in the last section.

\section{$3D$ black hole conformally coupled to a   scalar}\label{sec2}
In this section, we briefly study the properties of the metric of $ 3D $   black hole conformally coupled to a   scalar.
The action used in this theory, the field equations corresponding to metric and also the scalar field, are respectively as follows \cite{Martinez,Sudhanshu}
\begin{equation}\label{metric}
\mathcal{I} = \int {{d^3}x\,\sqrt { - g} \left[ {\mathcal{R} + \frac{2}{{{l^2}}} - {\nabla _\mu }\varphi {\nabla ^\mu }\varphi   - \frac{1}{16}\mathcal{R}{\varphi ^2}} \right]} ,
\end{equation}
where $\mathcal{R}$ is the Ricci scalar, $l$ is length scale which corresponds to the cosmological constant and  $\varphi$ is the
conformal scalar field.
The equations of motion are
\begin{equation}\label{field}
\mathcal{G}_{\mu \nu } - {g_{\mu \nu }}{l^{ - 2}} = {\nabla _\mu }\phi {\nabla _\nu }\phi  - \frac{1}{2}{g_{\mu \nu }}{\nabla ^\alpha }\phi {\nabla _\alpha }\phi  + \frac{1}{8}({g_{\mu \nu }}{\nabla ^\mu }{\nabla _\mu } - {\nabla _\mu }{\nabla _\nu } + {\mathcal{G}_{\mu \nu }}){\phi ^2},
\end{equation}\begin{equation}\label{scalar}
 {\nabla ^\mu }{\nabla _\mu }\phi - \frac{1}{8}\mathcal{R} \phi  = 0,
\end{equation}
where $ \mathcal{G}_{\mu \nu }= {\mathcal{R}_{\mu \nu }} - \frac{1}{2}{g_{\mu \nu }}\mathcal{R}$.

Now, the metric of the spherically symmetric spacetime can be written as
\begin{equation}\label{metric}
 d{s^2} =  - f(r)d{t^2} + f{(r)^{ - 1}}d{r^2} + {r^2}d{\theta ^2} ,
\end{equation}
where metric function has the form:
\begin{equation}
f(r) = \frac{{{{(r + b )}^2}(r - 2b)}}{{{l^2}r}},
\end{equation}
along with the matter field configuration \cite{Martinez}
\begin{equation}
\varphi =\sqrt{\frac{8b}{r+b}}.
\end{equation}
Here $b$ is the constant of integration.
The radius of the horizon ($ r_ h  $) satisfy the horizon condition $ {\left. f(r)\right|_{r = {r_h}}}=0$ and so, $ b=\dfrac{r_ h}{2} $.
The details of this solution can be found in Refs.  \cite{Martinez,Sudhanshu}.

\section{Thermodynamics}\label{sec3}

In this section, we focus on the thermodynamics of $3D$ black hole conformally coupled to a   scalar. In particular,
we study the thermodynamics in the perspective of non-perturbative (exponential) correction  attributed due to the black hole entropy under the consideration of quantum nature of black hole.
The Hawking temperature and the entropy density for this particular black hole are given as follows \cite{Sudhanshu}
\begin{equation}\label{S0}
 T = \frac{{9{r_h}}}{{16\pi {l^2}}} , \quad\quad {S_0} = \frac{{\pi {r_h}}}{3}.
\end{equation}
It is interesting to note that while $r_h\rightarrow0$ then $T\rightarrow0$, so it is completely opposite of the Schwarzchild black hole, where $r_h\rightarrow0$ yields $T\rightarrow\infty$, so the Schwarzchild black hole was unstable. However, we will show that the $3D$ black hole conformally coupled to a  scalar is stable in absence of the quantum corrections. It is well-known that the significance of perturbation can not ignored for the study of reduced size of the black hole considerably due to the
Hawking radiation.  In fact, for a thermodynamical system, the
  thermal fluctuations lead to the corrections to the ordinary entropy
by logarithmic term (at leading order) as a perturbative correction  (\cite{BM,BSSM,BSHH,BMZA,BAI})
and by exponential term as a non-perturbative quantum correction (\cite{BM,SYR,AA}).
In each of the cases, efforts have been made to study various issues of gravity, general relativity and quantum theory of gravity, etc.
that provide useful information appropriately.
The total entropy in terms of quantum corrections is as follows \cite{BM,BSSM,BSHH,BMZA,BAI,SYR,AA}
\begin{equation}
S = S_0 + \alpha \ln S_0 + \dfrac{\lambda}{S_0} + \eta {e^{ - S_0}} + \ldots ,\label{ex}
\end{equation}
where, $ \alpha $, $ \lambda $ and $ \eta $ are constants and $ S_0 $ is the original equilibrium entropy. However, the second and third terms in the R.H.S. of (\ref{ex}) are related to perturbative correction,
and the fourth term is related to non-perturbative quantum corrections.
Therefore, we can write the  expression for the  non-perturbatively corrected entropy as
 \begin{equation}
S = S_0  + \eta {e^{ - S_0}}. \label{non}
\end{equation}
The perturbative correction on the entropy has been discussed in Ref. \cite{Sudhanshu}.
Here, we try to extend the investigation by discussing the effect of the non-perturbative quantum correction on the stability of $3D$ black hole conformally coupled to a   scalar. Henceforth, utilizing relations in  (\ref{S0}), the  non-perturbative corrected entropy (\ref{non}) for this black hole leads to
\begin{equation}\label{SC}
S = \dfrac{\pi r_{h}}{3}+\eta {e^{ - \dfrac{\pi r_{h}}{3}}}  .
\end{equation}
The first-law  of thermodynamics with no work term is given by
\begin{eqnarray}
dE=TdS,
\end{eqnarray}
where $E$ is total mass (internal energy) of the system. 
However, in quantum gravity scenarios the first law, may receive some corrections \cite{R5}.\\

\begin{figure}[h]
	\centering
	\subfigure{
		\includegraphics[width=0.50\textwidth]{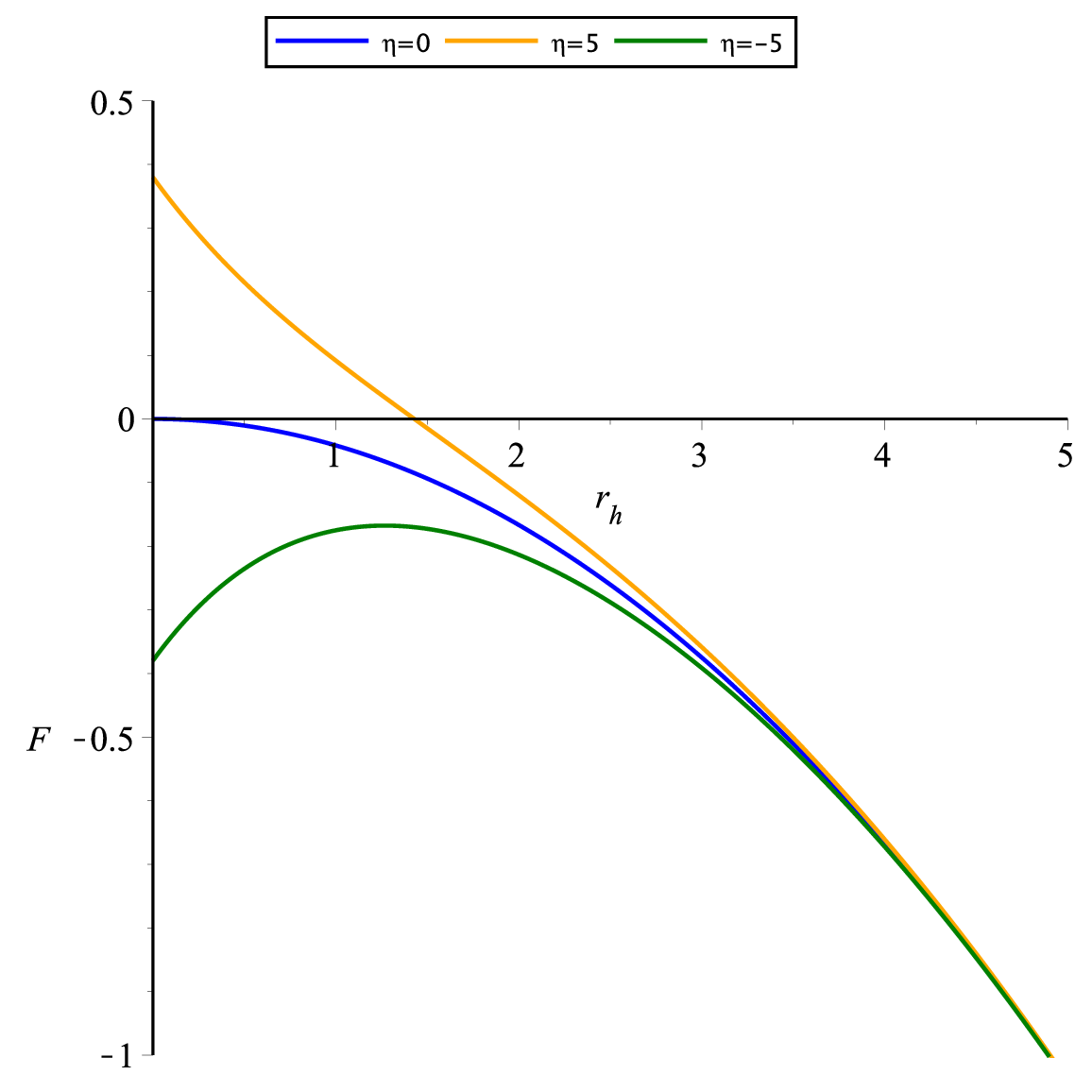}
}
	\caption{Variations of the Helmholtz free energy in terms of horizon radius $ r_{h} $.}
	\label{Pic:F}
\end{figure}

\begin{figure}[h]
 	\centering
 	\subfigure{
 		\includegraphics[width=0.50\textwidth]{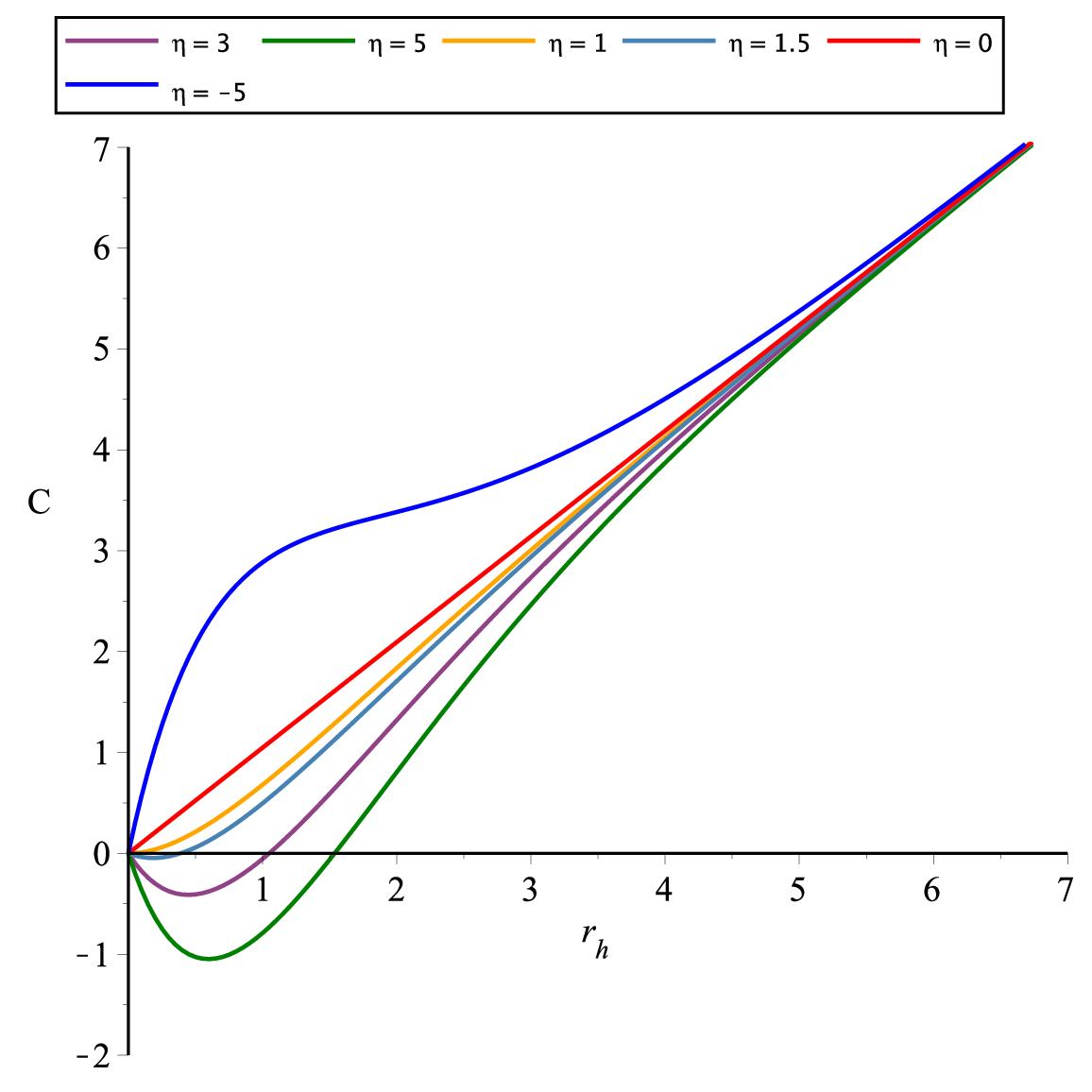}
 	}
 	\caption{Variations of the original and the corrected specific heat in terms of horizon radius $ r_{h} $.}
 	\label{Pic:C}
 \end{figure}

In the following, some important issues are considered in the investigation of the given black hole. In particular, we systematically examine the effects of quantum correction on the thermodynamical variables.  One of these cases is the investigation of Helmholtz free energy, which can be fruitful in analyzing black hole stability and phase transition.
The Helmholtz free energy can be written through the following general equation:
\begin{equation}\label{FA}
F =  - \int {SdT}
\end{equation}
Therefore, using the Eqs. (\ref{S0}), (\ref{SC}) and (\ref{FA}), we have
\begin{equation}\label{F}
F = - \frac{3}{{32}}  \frac{{{r_h}^2}}{{{l^2}}} +\frac{27}{{16}}  \frac{{   \eta    }}{{{\pi ^2}{l^2}}} {e }^{ - \frac{{\pi {r_h}}}{3}}.
\end{equation}
Here, we see that for  $ \eta=0 $, we have
\begin{equation}\label{F0}
F_{0}= - \frac{3}{{32}}{\mkern 1mu} \frac{{{r_h}^2}}{{{l^2}}}.
\end{equation}
In order to see the behavior of  the Helmholtz free energy, we plot Fig. \ref{Pic:F}.
Here, we observe that the exponential correction play an important role
for the Helmholtz free energy of the small sized black holes only. The values of the Helmholtz free energy are positive and negative for positive  and negative correction parameter ($\eta$), respectively. When the Helmholtz free energy is positive the black hole is in the unstable phase (and vise versa). It is clear by comparing Fig. \ref{Pic:F} and Fig. \ref{Pic:C}.\\

\begin{figure}[h]
	\centering
	\subfigure{
		\includegraphics[width=0.50\textwidth]{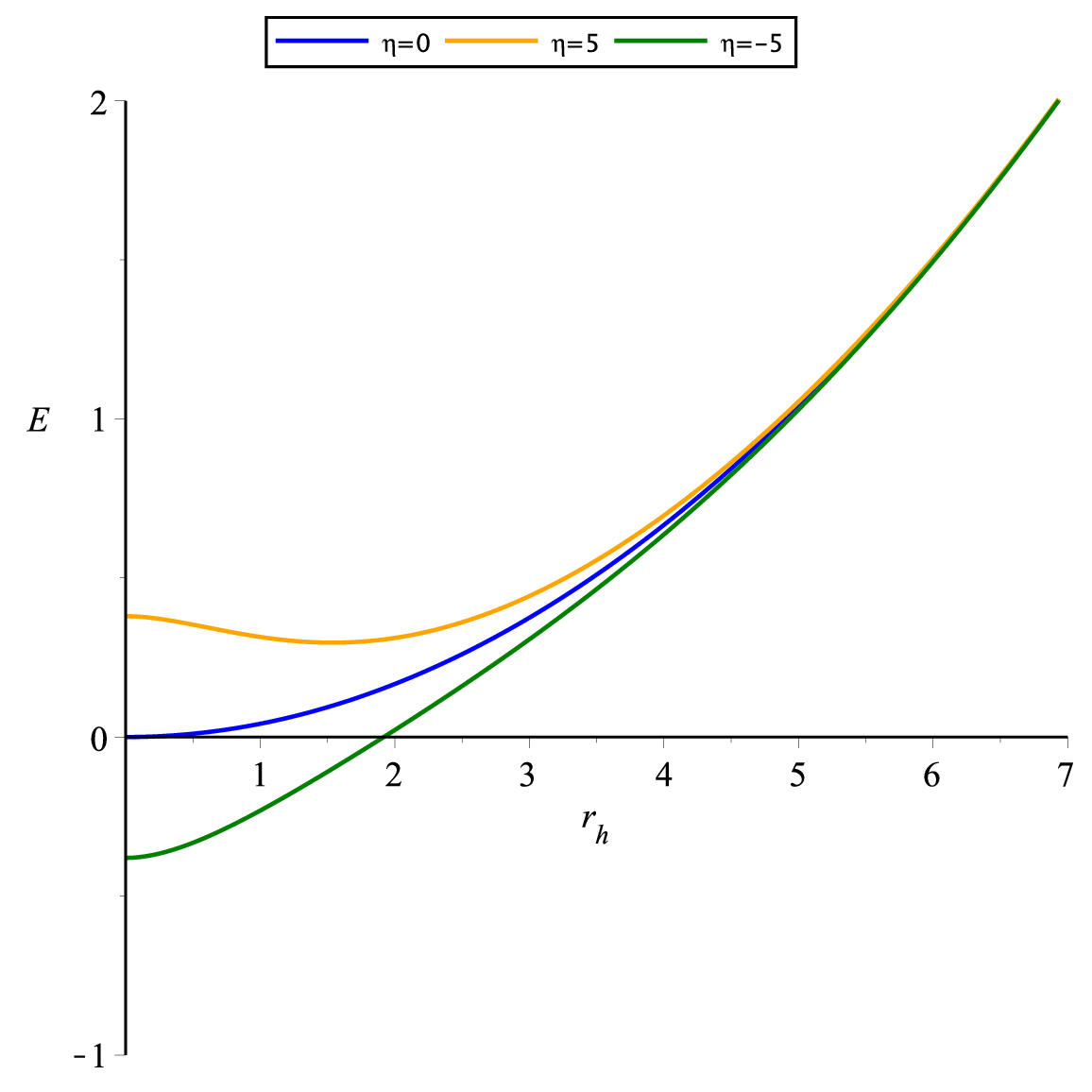}
}
	\caption{Variations of the internal energy in terms of horizon radius $ r_{h} $.}
	\label{Pic:E}
\end{figure}

Another important quantity for thermodynamics is the internal energy. We compute it under the effect of exponential quantum correction.
The general expression for the internal energy of the thermodynamic system is  given by
\begin{equation}\label{E0}
E = \int {TdS}.
\end{equation}
So, using the Eqs. (\ref{S0}), (\ref{SC}) and (\ref{E0}) we have
\begin{equation}\label{E}
E = \frac{3}{{32}} \frac{{{\pi ^2}{r_h}^2 + 6  \pi  {r_h}\eta {\mkern 1mu} {{\rm{e}}^{ - \frac{{\pi   {r_h}}}{3}}} + 18  \eta  {{ {e}}^{ - \frac{{\pi   {r_h}}}{3}}}}}{{{\pi ^2}{l^2}}}.
\end{equation}
Moreover, the equilibrium value of the internal energy  (for $ \eta=0 $) is obtained by
\begin{equation}
E_{0}=\frac{3}{{32}}{\mkern 1mu} \frac{{{r_h}^2}}{{{l^2}}}.
\end{equation}
This confirms that the equilibrium value of internal energy is equal to negative of Helmholtz free energy.
Now, we plot   the internal energy with respect to the horizon radius as shown  in Fig. \ref{Pic:E}.
From the plot, we see that the exponential correction is responsible for internal energy to have non-zero value when   black holes ceases to the point sized black holes.
Similar to the Helmholtz free energy, the internal energy takes
 negative and positive values for negative and positive correction parameter, respectively. \\
 
 Moreover, the corrected specific heat for this system can be written as
 \begin{equation}\label{C}
 	C=\frac{dE}{dT}=- \dfrac{\pi r_{h}}{3}\left(\eta{e^{ - \frac{\pi}{ r_{h}}{3}}}-1\right).
 \end{equation}
 In addition, the original specific heat $ C_{0} $ for the case ($ \eta  = 0 $), is as follows
 \begin{equation}\label{C0}
 	{C_0} =  \dfrac{\pi r_{h}}{3}.
 \end{equation}
 To have more information and to see the results of applied quantum correction, the heat capacity   in terms of horizon radius $ r_{h} $ is depicted in Fig. \ref{Pic:C}.
 From the  Fig. \ref{Pic:C}, we find that the heat capacity of this black hole for the initial state without
 quantum correction ($  \eta =0 $) is completely positive and has no phase transition.
 This state is also established for the application of quantum correction with the negative correction coefficient.
 However, for the quantum corrected specific heat with positive coefficient we can see a first order phase transition which may be interpreted as large/small black hole phase transition. It means that a large black hole is in the thermodynamically stable phase while a small black hole is unstable due to the quantum effects.
 
 \clearpage
 
\section{Quantum work}\label{sec04}
Black hole entropy, at quantum scales, is related to the microstates. Hence, a change in the black
hole microstates yields to the change in the entropy. So, one can express the change in the quantum
corrected entropy of the conformally dressed $3D$ black hole as,

\begin{equation}\label{Delta S}
\Delta S=S_{f}-S_{i},
\end{equation}
where $S_{f}$ is the final entropy, while $S_{i}$ is the initial entropy in an evolution. We consider the initial state of a large black hole with $r_{hi}$ (where quantum correction is negligible) which is yields to the unstable final state with the $r_{hf}$ hence using the equation (\ref{SC}), one can obtain,
\begin{equation}\label{Delta S2}
\Delta S=\dfrac{\pi (r_{hf}-r_{hi})}{3}+\eta {e^{ - \dfrac{\pi r_{hf}}{3}}}.
\end{equation}
In that case one can obtain the change in the Helmholtz free energy, which help us to study the amount of quantum work. Therefore, the change of Helmholtz free energy ($\Delta F=F(r_{hf})-F(r_{hi})$) can be expressed as
\begin{equation}\label{Delta-F}
\Delta F = - \frac{3}{{32}}  \frac{({{r_{hf}}^2}-{{r_{hi}}^2})}{{{l^2}}} +\frac{27}{{16}}  \frac{{   \eta    }}{{{\pi ^2}{l^2}}} {e }^{ - \frac{{\pi {r_{hf}}}}{3}}.
\end{equation}
Now, the quantum work ($e^{\frac{\Delta F}{T}}$) is expressed as \cite{QW},

\begin{equation}\label{W}
<e^{\frac{W}{T}}>= \exp\left(- \frac{\pi}{{6}}  \frac{({{r_{hf}}^2}-{{r_{hi}}^2})}{{{r_{h}}}}\right) \exp\left(  \frac{{   3\eta    }}{{{\pi}{r_{h}}}} {e }^{ - \frac{{\pi {r_{hf}}}}{3}}\right).
\end{equation}

In fact, one can imagine a dual picture of a black hole geometry in a superconformal field theory. The black geometry of a conformally dressed $3D$ black hole emits Hawking radiation. Its corresponding energy is described by heat, which is yields to
information loss paradox in quantum thermodynamics. Such an information loss paradox may be solved using a dual picture in a superconformal field theory, when we study the quantum behavior. It is indeed characteristic of the superconformal side in the mentioned duality. Therefore, a conformally dressed $3D$ black hole is represented by the unitary superconformal field theory. Hence, we are able to use the quantum non-equilibrium thermodynamics to study this black hole geometry. In that case we succeed to calculate quantum work of a conformally dressed $3D$ black hole. This quantum work is denoted by an unitary information
preserving process which is the dual picture of a conformally dressed $3D$ black hole. Hence, in the other side of duality we have a non-equilibrium quantum thermodynamics. The non-equilibrium quantum thermodynamics enables a conformally dressed $3D$ black hole to lose its mass through an unitary process.  We have shown that a conformally dressed $3D$ black hole is unstable through a first order phase transition and will evaporate at the final state. So information can leak out of a conformally dressed $3D$ black hole during the last stages of its evaporation from the mentioned unitary information preserving process.
Therefore, we can resolve the information loss paradox in the given system
using the non-equilibrium quantum thermodynamics. It is clear that at the quantum scale, we need
to analyze a given system using the non-equilibrium quantum thermodynamics.

\section{Thermodynamic geometry}\label{sec4}
\begin{figure}[h]
	\centering
	\subfigure{
		\includegraphics[width=0.50\textwidth]{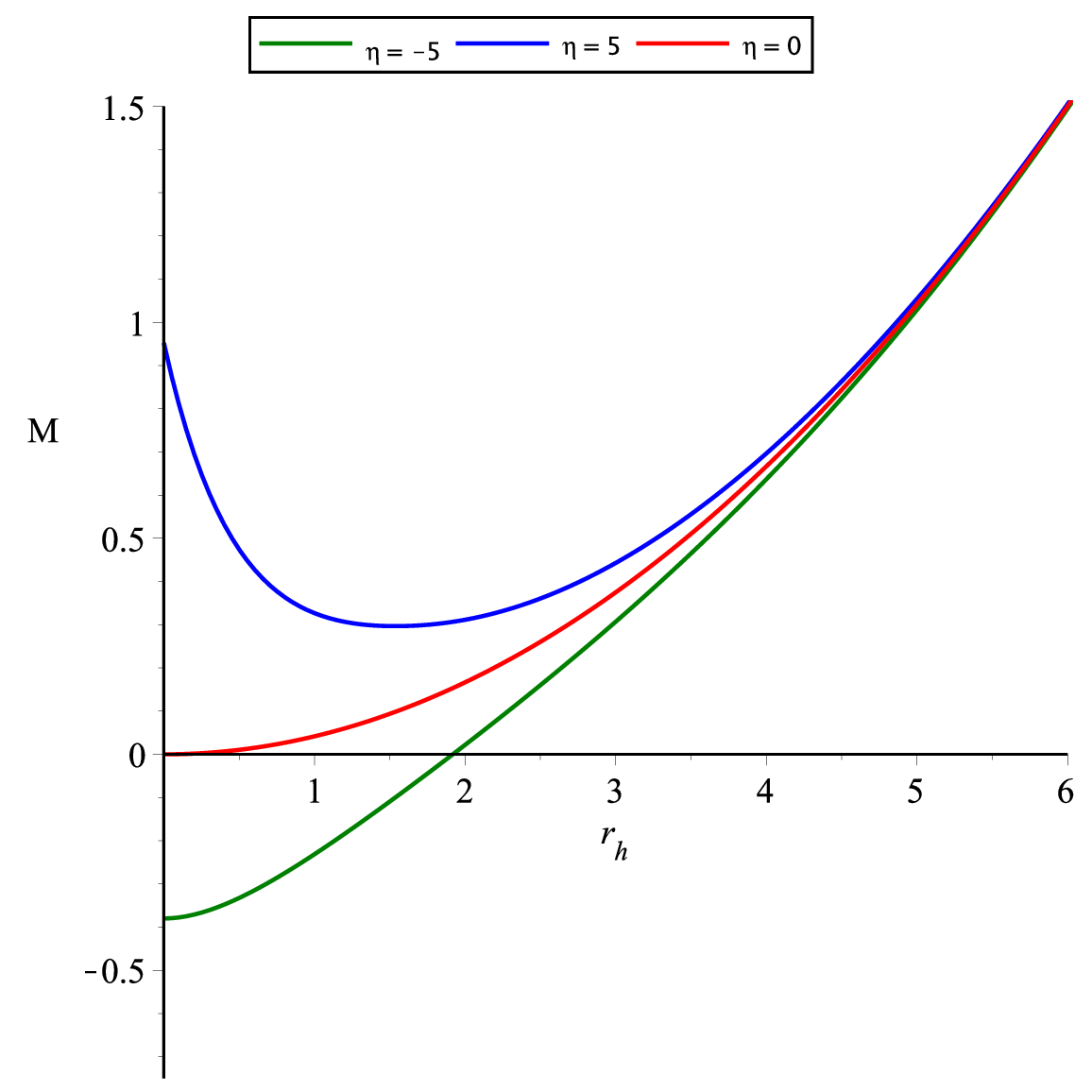}
}
	\caption{Variations of mass in terms of horizon radius $ r_{h} $.}
	\label{Pic:M}
	
\end{figure}

\begin{figure}[h]
	\centering
	\subfigure{
		\includegraphics[width=0.50\textwidth]{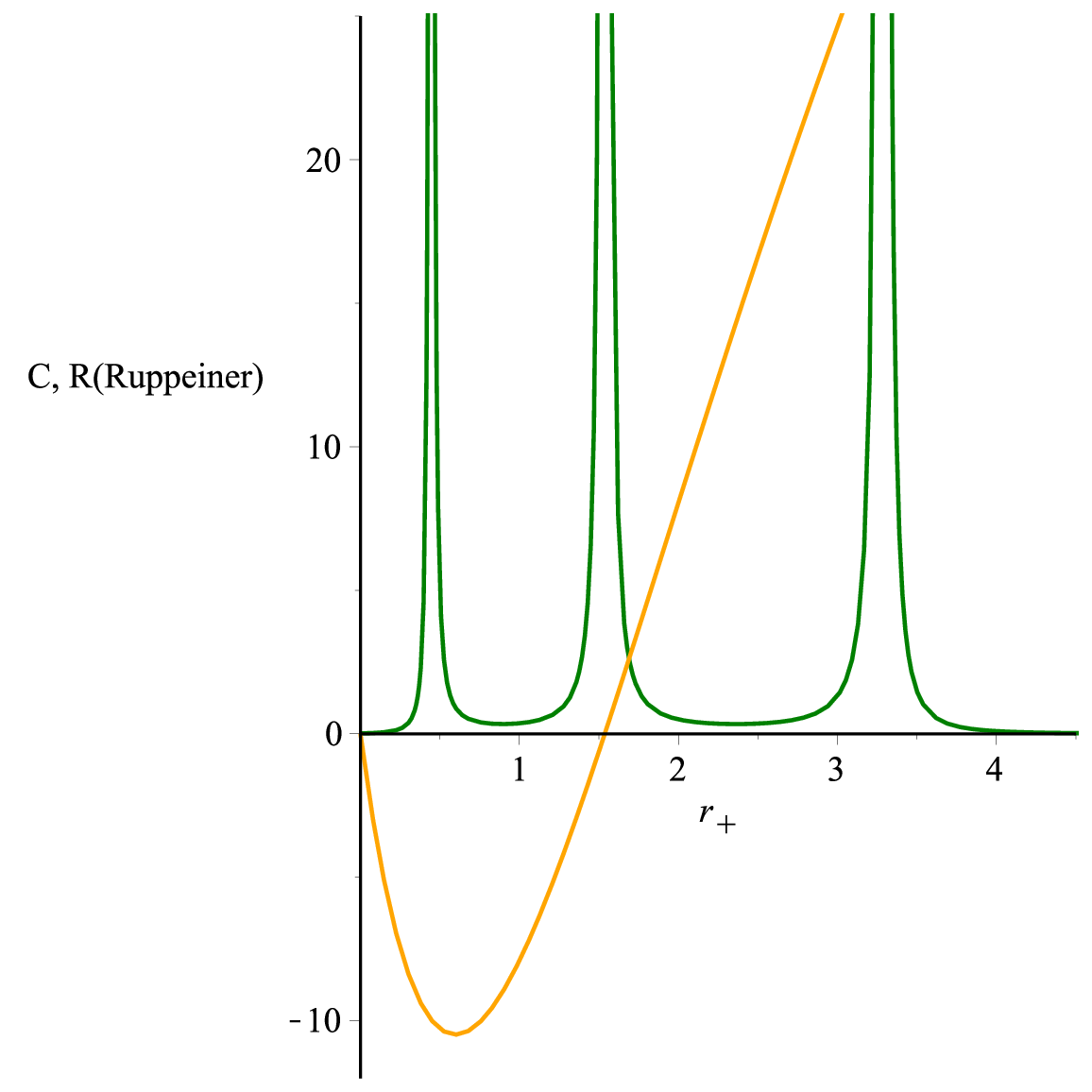}
}
	\caption{Variations of the Ricci scalar of Ruppiner formalism and heat capacity in terms of horizon radius $ r_{h} $.}
	\label{Pic:CR}
\end{figure}

Another method to investigate the stability of black holes is thermodynamic geometry.
In other words, the phase transition of the black hole can be investigated by using divergences of the Ricci scalar \cite{Dimov,Vetsov,Sheykhi,Li,Wei,Wei2,Dehghani,Dehghani2,Soroushfar}.
Here, by including quantum correction, we study the geometric structure for a $3 D$  black hole conformally
coupled to a  scalar, according to geometric formalism suggested by Ruppiner \cite{Ruppeiner,Ruppeiner1}.
The quantum corrected effective Ruppeiner metric can be written as \cite{Ruppeiner,Ruppeiner1,Pourhassan,Pourhassan1}
\begin{equation}\label{Rup}
d{s^2} =  - \frac{1}{T}Mg_{ab}^Rd{X^a}d{X^b}.
\end{equation}
Taking  the quantum correction and $ M \equiv E\ $ into account,  the mass of the black hole can be written as follows
\begin{equation}\label{M}
M(S,l) =  - \frac{{27}}{{32}} \frac{{{{\left( { W\left( { - \eta  {{\rm{e}}^{ - S}}} \right)} \right)}^2} - {S^2} + 2   W\left( { - \eta   {{ {e}}^{ - S}}} \right)}}{{{\pi ^2}{l^2}}},
\end{equation}
where $W$ is the Lambert $W$ function.
Plot of mass is shown in Fig.~\ref{Pic:M}.  From the plot, one also confirms that  the
mass shows same behavior as the internal energy as justified by   the thermodynamic geometry.
Also, plot of the Ricci scalar of Ruppiner formalism comparing to heat capacity is demonstrated in Fig.~\ref{Pic:CR}.


As one can see from this figure that the Ricci scalar of Ruppiner metric has singularities, in
which one of them is exactly on the zero heat capacity of the black hole in the quantum correction state.
In other words, the divergence of the scalar curvature of Ruppiner formalism corresponds to the phase transition point of this black hole.

\section{Conclusion}\label{sec5}
In this paper, we have considered the $3D$ black hole conformally coupled with a scalar field to study the effects of non-perturbative quantum correction to  their thermodynamics.  In particular,  we have calculated the
exponentially corrected entropy which attributes the correction to the other thermodynamical variables. The heat capacity is calculated from the first-law of thermodynamics and found that the heat capacity for the initial state without
quantum correction is completely positive and has no phase transition. This state is
also established for the  quantum correction with the negative correction parameter. However, due to quantum correction with the positive correction parameter, we see that
the heat capacity of this black hole has a zero point and has a first order phase transition. The small black holes are unstable for positive correction parameter and   correction terms with positive correction parameter do not
change the state of black hole. Furthermore, we have calculated the Helmholtz free energy and internal energy under the influence of
non-perturbative correction. Here, we find that the exponential correction
plays a significant role to the thermodynamics of small sized black holes. We also calculated the quantum work of conformally dressed $3D$ black hole. The quantum work is also discussed for the given black hole.

Furthermore, we have  studied  the  stability and phase transition from the perspectives of thermodynamic geometry. Here, we have calculated the mass of the black hole appeared in the quantum corrected effective Ruppeiner metric.
The Ricci scalar of Ruppiner metric possesses singularities.
This confirmed that the divergence of the scalar curvature of Ruppiner formalism corresponds to the phase transition point of this black hole.

Moreover, this method and the proposed analysis can be used to calculate the entropy corrections on the scale possessing black holes of a similar theory such as \cite{Chan:1996rd,Tang:2019jkn,Karakasis:2021ttn}.

\end{document}